\documentstyle[preprint,pra,aps]{revtex}

\begin{document}

\draft

\tightenlines

\title{Duality in Perturbation Theory and the Quantum Adiabatic Approximation}

\author{Marco Frasca}
\address{Via Erasmo Gattamelata, 3,
         00176 Roma (Italy)}

\date{\today}

\maketitle

\abstract{Duality is considered for the
perturbation theory by deriving, given a series solution in a
small parameter, its dual series with the development parameter
being the inverse of the other.
A dual symmetry in perturbation theory is identified. It is
then shown that the dual to the Dyson series in quantum mechanics
is given by a
recently devised series having the adiabatic approximation as
leading order. A simple application of this result is given
by rederiving a theorem for strongly perturbed quantum systems.}


\narrowtext

A known result of fluid mechanics \cite{pnueli}, given the
Navier-Stokes equation in absence of any forcing
\begin{equation}
    \frac{\partial {\bf u}}{\partial t}+({\bf u}\cdot\nabla){\bf u}
    -\nu\nabla^2{\bf u}+\nabla p=0
\end{equation}
where $\nu$ is the viscosity, $p$ the pressure and ${\bf u}$ the
velocity field, is that, by fixing the 0-th order solution through the
Eulerian part as \cite{sing}
\begin{equation}
    \frac{\partial {\bf u}_0}{\partial t}+({\bf u_0}\cdot\nabla){\bf u_0}
    =0,
\end{equation}
one obtains a perturbation series for a large Reynolds number $Re$
while, taking at the leading order the equation
\begin{equation}
    \frac{\partial {\bf w}_0}{\partial t}
    -\nu\nabla^2{\bf w}_0=0
\end{equation}
one obtains a perturbation series for small $Re$. So far, this was the only
case in perturbation theory, as applied to physics, with an equation
generating both a small perturbation series and its strong
perturbation counterpart.

Duality in perturbation theory should then be understood in the sense that,
for a given differential equation, one has the
possibility to derive perturbation
series both in a small development parameter and its inverse,
giving in this way the possibility to study a solution in
different regions of the parameter space.
However, it should be said that the method could not
be of absolutely general usefulness as some limitations can appear
for the computation both of the leading and higher orders. Beside, there
exists situations where better approximations are known, as I will show.
However, it is easy to realize that a lot of problems in physics can get new
insights from this approach, so that, it is worthwhile to be
exploited.

A natural question, in the light of the above defined duality in
perturbation theory, is what should be the dual to the well-known
Dyson series for the Schr\"{o}dinger equation. The answer to this
question is the main aim of this communication. In fact, the existence of
this possibility gives a new technique to analyse quantum
systems in different regions of their parameter space.

Quite recently, I showed that new solutions for the Schr\"{o}dinger
equation, in time-dependent problems, can be obtained when a strong
perturbation is applied to a quantum system \cite{Fra1}. This
approach seems to indicate that, at the leading order, an adiabatic
approximation should be used\cite{Fra2}. On a different line of
research, Mostafazadeh \cite{Most} was able to show that a series exists,
for the Schr\"{o}dinger equation, with a well
defined development parameter, having the adiabatic approximation
as leading order. Using duality in perturbation theory the above
different research lines can be merged, as I am going to show, giving
us the main result of this paper. In fact, the series derived by
Mostafazadeh is dual to the Dyson series. Then, the
theory of strong perturbations in quantum mechanics can be proved
to be dual to the standard small perturbation theory.

In order to show how a dual series can be obtained
in a simple case, let us consider
the model given by the following differential equation
\begin{equation}
    \ddot{x} = f_0(x)+\lambda f_1(x) \label{eq:main}
\end{equation}
where the dots mean derivation with respect to the time and
$\lambda$ is an ordering parameter. It is a well-known
matter that, when $\lambda\rightarrow 0$, a solution series of the form
$x\sim x_0 + \lambda x_1 + \lambda^2 x_2 + O(\lambda^3)$
can be obtained. However,
as for the Navier-Stokes equations, we are free to choose at the leading
order, as an unperturbed equation, $\ddot{x}_0=\lambda f_1(x_0)$. To
show that this choice gives a dual perturbation series,
I rescale the time variable in
eq.(\ref{eq:main}) as $t\rightarrow \sqrt{\lambda}t=\tau$. One gets
\begin{equation}
    \lambda\ddot{x} = f_0(x)+\lambda f_1(x) \label{eq:resc}
\end{equation}
where now the dots mean derivation with respect to $\tau$.
It is quite easy to verify that the series
\begin{equation}
    x = x_0 + \frac{1}{\lambda}x_1+\frac{1}{\lambda^2}x_2
    +O\left(\frac{1}{\lambda^3}\right)
\end{equation}
is a solution of eq.(\ref{eq:resc}) when
\begin{eqnarray}
    \ddot{x_0} &=& f_1(x_0) \nonumber \\
    \ddot{x_1} &=& f_1'(x_0)x_1 + f_0(x_0) \label{eq:terms} \\
    \ddot{x_2} &=& f_0'(x_0)x_1 + f_1'(x_0)x_2 +
    \frac{1}{2} f_2''(x_0)x_1^2 \nonumber \\
              &\vdots&  \nonumber
\end{eqnarray}
By analogy with the results in quantum mechanics \cite{Fra1},
I take the above as the dual method to small perturbation theory to
obtain a dual perturbation series to a given one. It is important to
note that the above result is true independently of one's ability
to solve the leading order equations.

We see that the arbitrariness in the choice of the leading order
equation gives rise to a symmetry. In fact, putting $\lambda=1$
into eq.(\ref{eq:main}), there is no more reason to
see any difference between the perturbation and the unperturbed
system in the same way as happens in fluid mechanics. This means
that the series given by the small perturbation theory can be derived
from the one given by the dual method and vice versa by simply
interchanging $f_0(x)\leftrightarrow f_1(x)$.
That is a symmetry of the perturbation theory whose meaning
can be really understood
only after the introduction of the dual series. Actually, the
general solution of eq.(\ref{eq:main}) can be written, for the
one-dimensional case, as
\begin{equation}
    t-t_0=\int_{x_0}^x dx'
    \frac{1}{\sqrt{2}\sqrt{E+\int_{x_0}^{x'}f_0(x'')dx''
    +\int_{x_0}^{x'}f_1(x'')dx''}}
\end{equation}
with $E$ a motion constant. It is easily seen that both the series
expansions, for small $f_0$ or $f_1$, can be straightforwardly
obtained. What it is interesting is that, the small parameter in a case
is the inverse of the development parameter in the other.
From the discussion above it
should be clear that both series can have the same problems as
secularities or divergent terms.

A more interesting example is given by the following Duffing equation
\begin{equation}
    \ddot{x} + \omega_0^2 x + \beta x^3 = f_0\cos(\omega t).
\end{equation}
By setting $\tau=\omega_0 t$, $\nu=\frac{\omega}{\omega_0}$,
$\xi=\frac{\omega_0^2}{f_0}x$ and
$\lambda=\frac{\beta f_0^2}{\omega_0^6}$, one gets the rescaled equation
\begin{equation}
    \ddot{\xi} + \xi + \lambda \xi^3 =
    \cos(\nu \tau). \label{eq:Duff}
\end{equation}
where the dots mean
derivation with respect to $\tau$ and  $\lambda$ is
just a parameter measuring the strength of the nonlinearity.
Eq.(\ref{eq:Duff}) is generally considered, analytically,
only for small $\lambda$, but what
happens in the limit of a very strong nonlinearity? Duality
can be applied and one easily realizes that, for large values of the
parameter $\lambda$, the quantity $\epsilon=\frac{1}{2}\dot{\xi}^2+
\lambda\frac{1}{4}\xi^4$ tends to be a constant of motion. This is due to
the result that the perturbation completely drives the system. So, we
have regular periodic motion in the considered limit. This
example shows that, although the leading order equation can be easy to solve,
going to higher orders could be very involved.

A class of important problems arises from the Schr\"{o}dinger
equation that I consider in the one-dimensional form
\begin{equation}
    -\frac{\hbar^2}{2m}\frac{d^2\psi}{dx^2} +
    V_0(x)\psi + \lambda V_1(x)\psi=E\psi
\end{equation}
where $\lambda\rightarrow\infty$. One could apply immediately the symmetry
between the dual and small perturbation theories discussed so far and
use without difficulty the Rayleigh-Schr\"{o}dinger approximation scheme.
While that is a correct approach, I will show how the dual
method works in this case. So, let us put $\xi=\sqrt{\lambda}x$,
$\psi=\psi_0+\frac{1}{\lambda}\psi_1+O\left(\frac{1}{\lambda^2}\right)$ and
$E=\lambda E_0 + E_1 + O\left(\frac{1}{\lambda}\right)$.
This yields the following equations
\begin{eqnarray}
    -\frac{\hbar^2}{2m}\frac{d^2\psi_0}{d\xi^2} + V_1(\epsilon\xi)\psi_0
    &=& E_0 \psi_0 \nonumber \\
    -\frac{\hbar^2}{2m}\frac{d^2\psi_1}{d\xi^2} + V_1(\epsilon\xi)\psi_1
    + V_0\psi_0    &=& E_1 \psi_0 + E_0 \psi_1\\
              &\vdots&  \nonumber
\end{eqnarray}
where $\epsilon=\frac{1}{\sqrt{\lambda}}$. At the leading order we get a
well-known equation, that is, a second order differential equation with
a slowly varying coefficient due to the perturbation.
In this case we can apply the WKB approximation
\cite{Nay}. Thus, the dual method yields in this case
a solution that is a combination of both Rayleigh-Schr\"{o}dinger and
semiclassical methods.

There are several problems where the above approximation can be
applied. A well-known example is given by the anharmonic oscillator that
has a large body of literature \cite{anh} and is a model which any
approximation scheme should address. The Hamiltonian can be cast
in the form
\begin{equation}
    H = \frac{p^2}{2}+\frac{1}{2}q^2+\frac{\lambda}{4}q^4. \label{eq:anh}
\end{equation}
The method I discussed so far gives an unambigous answer to this problem,
i.e. the leading order approximation, when the anharmonicity is very
strong, can be obtained by solving the equation
\begin{equation}
    \left(\frac{p^2}{2}+\frac{\lambda}{4}q^4\right)\psi_0(q)=E\psi_0(q).
\end{equation}
The quartic oscillator is well-known in the literature \cite{Parisi}.
So, we can compare
our method with numerical results. To leading order of the WKB
approximation of the energy levels, normalized in unit of
$\left(\frac{\hbar^2}{2}\sqrt{\frac{\lambda}{4}}\right)^{\frac{2}{3}}$,
the agreement is within 18\% with the true value of the ground state
energy for the anharmonic oscillator.
That agreement improves for higher excited states. However, we
know from Symanzick scaling that the quartic oscillator is the right
approximation for energy levels of the anharmonic oscillator
when $\lambda\rightarrow\infty$\cite{Simon}. Often, the use of
semiclassical eigenfunctions can be too much involved and
better approximation schemes, as those given in ref.\cite{anh},
can improve the situation. Duality, as applied in perturbation theory,
yields anyway a definite answer.

The situation is surely more interesting in time-dependent problems.
By noting that the only
meaningful quantities are transition probabilities between states of
the unperturbed system, we have the initial conditions definitely fixed
breaking in this way the dual symmetry of the perturbation theory.
In fact, in ref.\cite{Fra1} I showed that the problem
\begin{equation}
    (H_0+\lambda V(t))|\psi\rangle=i\hbar\partial_t|\psi\rangle
\end{equation}
with ${\displaystyle \partial_t=\frac{\partial}{\partial t}}$ and
$\lambda\rightarrow\infty$, using the above dual method, has
the leading order solution
\begin{equation}
    |\psi\rangle \sim U(t)|n\rangle + O\left(\frac{1}{\lambda}\right)
\end{equation}
with
\begin{equation}
    \lambda V(t) U(t) = i\hbar\partial_t U(t)
\end{equation}
and $H_0|n\rangle =E_n|n\rangle$.
So, the unperturbed solution fixes the initial condition
as also happens in the small perturbation theory.
But, in order to leave the dual simmetry untouched, one should
physically consider also systems initially prepared with the
eigenstates of the perturbation, but this is not
the case for the computation of probability transitions.
Then, we can conclude that,
for the time-dependent perturbation theory as usually applied in
quantum mechanics, the dual symmetry is broken due to the choice
of the initial conditions.

By the methods discussed above, we can obtain the main
result of the paper. Our aim is to show that the dual
to the Dyson perturbation series,
for the time-dependent Schr\"{o}dinger equation, is given
by the series obtained by Mostafazadeh \cite{Most} having
the adiabatic approximation as leading order. So, let us
consider the Schr\"{o}dinger equation $H(t)U=i\hbar\partial_t U$
being U the time evolution operator. The Dyson series is the
solution of that equation and can be written formally as
\begin{equation}
    U(t)={\cal T}\exp{\left(-\frac{i}{\hbar}\int_0^tdt'H(t')\right)}
\end{equation}
with ${\cal T}$ the time-ordering operator.
This is a compact writing for the series development
\begin{equation}
    U(t)=I-\frac{i}{\hbar}\int_0^tdt'H(t')
          +\left(-\frac{i}{\hbar}\right)^2
          \int_0^tdt'\int_0^{t'}dt''H(t')H(t'')+\cdots
\end{equation}
A dual series to the one above is meant as a series having for
development parameter its inverse, as discussed above. To obtain
it, let us consider the case with
the hamiltonian $H$ being constant in time. Assuming, for the
sake of simplicity, here and in the following
that the hamiltonian has a discrete spectrum, the solution to the
time-dependent Schr\"{o}dinger equation is easily obtained through
the time evolution operator
$U(t)=\sum_n e^{-\frac{i}{\hbar}E_nt}|n\rangle\langle n|$
with $H|n\rangle =E_n|n\rangle$, where
the effect of the time derivative is simply
$\partial_t U(t)=\sum_n \left(-\frac{i}{\hbar}E_n\right)
e^{-\frac{i}{\hbar}E_nt}|n\rangle\langle n|$. Instead, for
a time-dependent hamiltonian $H(t)$,
in the case of the adiabatic approximation
we have $U_A(t)=\sum_n e^{i\alpha_n(t)}|n,t\rangle\langle n,0|$ being
$\alpha_n(t)=\gamma_n(t)-\frac{1}{\hbar}\int_0^tdt'E_n(t')$
with $H(t)|n,t\rangle=E_n(t)|n,t\rangle$,
so that $E_n(t)$ gives the dynamical part
of the phase $\alpha_n(t)$ and
$\gamma_n(t)=i\int_0^tdt'\langle n,t'|\partial_{t'}|n,t'\rangle$
the geometrical part.
It is natural to ask how one can define
a derivative $D_t$ to obtain the same result as for the
time-independent case, that is,
$D_t U_A(t)=\sum_n \left(-\frac{i}{\hbar}E_n(t)\right)
e^{i\alpha_n(t)}|n,t\rangle\langle n,0|$.
It is quite easy to verify that the
following definition of $D_t$ has the requested property
\begin{equation}
    D_t=\partial_t+i\sum_{\stackrel{n,m}{n\neq m}}
    \langle m,t|i\partial_t|n,t\rangle|m,t\rangle\langle n,t|
\end{equation}
so that, the adiabatic approximation is exact for the equation
\begin{equation}
    H(t)U_A(t)=i\hbar D_tU_A(t)    \label{eq:adia}
\end{equation}
and it is easily verified that
$iD_t|k,t\rangle =\dot{\gamma}_k(t)|k,t\rangle$.
In this way, we are a step away from the sought result. In fact, let
us introduce a generic perturbation $V(t)$ into eq.(\ref{eq:adia}), so
that
\begin{equation}
    (H(t)+V(t))U(t)=i\hbar D_tU(t).
\end{equation}
In this form the duality principle of perturbation theory can be
applied. Having $V(t)$ as the unperturbed hamiltonian and
$H(t)$ as the perturbation, we get
no physical meaningful results from the leading order equation
$V(t)U^{(0)}(t)=i\hbar D_tU^{(0)}(t)$ unless we choose
$V(t)=H(t)$ or ${\displaystyle V(t)=-\sum_{\stackrel{n,m}{n\neq m}}
\langle m,t|i\hbar\partial_t|n,t\rangle|m,t\rangle\langle n,t|}$.
The latter is the interesting
case giving trivially the standard Dyson series.
So, when we take $H(t)$ as the unperturbed hamiltonian
and $V(t)$ as the perturbation, we have at the leading order
\begin{equation}
    H(t)U^{(0)}(t)=i\hbar D_tU^{(0)}(t),
\end{equation}
but this is nothing else than eq.(\ref{eq:adia}) and then
$U^{(0)}(t)=U_A(t)$ (i.e. at the leading order we have the
adiabatic approximation). To complete the identification with
the Mostafazadeh result, we have that the higher order
corrections are computed solving the equation
\begin{equation}
    H'(t)U'(t)=i\hbar \partial_tU'(t),
\end{equation}
where
\begin{equation}
    H'(t)=U_A^\dagger(t)V(t)U_A(t)=
    -\sum_{n,m,n\neq m}e^{-i(\alpha_m(t)-\alpha_n(t))}
    \langle m,t|i\hbar\partial_t|n,t\rangle|m,0\rangle\langle n,0|
\end{equation}
as given in ref.\cite{Most}. Then, one obtains \cite{Most}
\begin{equation}
    U(t)=U_A(t){\cal T}\exp{\left(-\frac{i}{\hbar}\int_0^tdt'H'(t')\right)}
\end{equation}
that completes the proof. No adiabatic hypothesis entered into this
argument as it should be.

As an application of that result, a theorem recently derived by me and,
in a rigourous way but in a different context,
by Joye \cite{Fra2} can be obtained for the
theory of the strong perturbations in quantum mechanics \cite{Fra1}.
In fact, it was proved that, for a quantum system described by
the Schr\"{o}dinger equation
$(H_0+\lambda V(t))|\psi\rangle=i\hbar\partial_t|\psi\rangle$,
in the limit
$\lambda\rightarrow\infty$, the adiabatic approximation, using the
eigenstates of the perturbation $V(t)$,
is a good approximation for $|\psi\rangle$. So, by
considering the perturbed system in the interaction picture,
gives the hamiltonian $H_I(t)=U^{(0)\dagger}(t)\lambda
V(t)U^{(0)}(t)$ where $H_0U^{(0)}(t)=i\hbar\partial_tU^{(0)}(t)$.
Then, the result obtained for the Dyson series by
duality can be directly applied to $H_I(t)$. We obtain
for a small $\lambda$ (otherwise we miss convergence), the
Dyson series
\begin{equation}
    U_I(t)=I-\frac{i}{\hbar}\int_0^tdt'H_I(t')
          +\left(-\frac{i}{\hbar}\right)^2
          \int_0^tdt'\int_0^{t'}dt''H_I(t')H_I(t'')+\cdots
\end{equation}
and, for a large $\lambda$, the Mostafazadeh result applied to
$H_I(t)$ having at the leading order the adiabatic approximation as
it should be. It must be noticed that, in the latter case, the eigenstates
to be considered are those of the perturbation. In fact, we
have $U^{(0)\dagger}(t)V(t)U^{(0)}(t)|k,t\rangle_I=
v_k^I(t)|k,t\rangle_I$ but
this is equivalent to
$V(t)(U^{(0)}(t)|k,t\rangle_I)=v_k^I(t)(U^{(0)}(t)|k,t\rangle_I)$.
Then, $v_k^I$ is an eigenvalue of $V(t)$
and $U^{(0)}(t)|k,t\rangle_I$ differs just
by a time-dependent phase factor from the corresponding
eigenstate of the perturbation.
As a by product, we get the confirmation that higher
order corrections are those computed by the method of
strong perturbation theory as given in \cite{Fra2}.

In summary I introduced the duality principle in perturbation
theory for differential equations.
A dual method with respect to the theory of
small perturbation is yielded and a
dual symmetry between the two methods
arises from the freedom in the choice of what the perturbation is.
The use of duality shows that, for the time-dependent
Schr\"{o}dinger equation, the dual series to the Dyson one is
given by a perturbation series computed recently by Mostafazadeh,
with a well-defined development parameter, having the
adiabatic approximation as leading order. This enriches the possibility
to analyze quantum systems in completely different regions of
their parameter space.

\end{document}